\documentclass[aps,prd,reprint,longbibliography,titlepage,nofootinbib]{revtex4-1} %,

\usepackage{amsthm}
\usepackage{amssymb}   
\usepackage{mathtools} 
\usepackage{hyperref}
\hypersetup{colorlinks=true,linkcolor=blue,urlcolor=blue,citecolor=blue}
\usepackage{accents}
\usepackage{tensor}
\usepackage[cal=boondox]{mathalfa}
\usepackage{lipsum}
\usepackage{relsize}
\usepackage{color}
\usepackage{comment}
\usepackage{nameref}
\usepackage[T1]{fontenc}
\newcommand{\ovg}[1]{\stackrel{(\gamma)}{#1}}
\newcommand{\uac}[1]{\underaccent{\tilde}{#1}}

\begin{document}
	
	\title{Canonical analysis with no second-class constraints of $BF$ gravity with Immirzi parameter}

	\author{Merced Montesinos\href{https://orcid.org/0000-0002-4936-9170} {\includegraphics[scale=0.05]{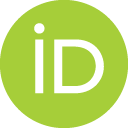}}}
	\email{merced@fis.cinvestav.mx}
	\affiliation{Departamento de F\'{i}sica, Cinvestav, Avenida Instituto Polit\'{e}cnico Nacional 2508,\\
	San Pedro Zacatenco, 07360 Gustavo A. Madero, Ciudad de M\'exico, Mexico}
	
	\author{Mariano Celada\href{https://orcid.org/0000-0002-3519-4736} {\includegraphics[scale=0.05]{ORCIDiD_icon128x128.png}}}
	\email{mcelada@matmor.unam.mx}
	\affiliation{Centro de Ciencias Matem\'{a}ticas, Universidad Nacional Aut\'{o}noma de M\'{e}xico,\\
	UNAM-Campus Morelia, Apartado Postal 61-3, Morelia, Michoac\'{a}n 58090, Mexico}
	
	\date{\today}
	
	\begin{abstract}
		In this paper we revisit the canonical analysis of $BF$ gravity with the Immirzi parameter and a cosmological constant. By examining the constraint on the $B$ field, we realize that the analysis can be performed in a Lorentz-covariant fashion while utterly avoiding the introduction of second-class constraints during the whole process. Finally, we make contact with the description of the phase space of first-order general relativity in terms of canonical variables with manifest Lorentz covariance subject to first-class constraints only recently introduced.
	\end{abstract}
	
	\maketitle
	
	%%%%%%%%%%%%%%%%%%%
	\section{\label{sec:Intro} Introduction}
	
	It is a well-established fact that Einstein's theory of general relativity can be expressed as a constrained $BF$ theory, something first materialized in Plebanski's formulation~\cite{pleb1977118} more than 40 years ago. In this sort of formulation, the gravitational field is encoded in a gauge connection and a 2-form on which a set of constraints must be imposed to break the topological character of the underlying $BF$ theory~\cite{cqgrevBF}. In turn, these formulations have served as the starting point of the so-called spinfoam models for quantum gravity~\cite{perez2013-16-3,rovelli2014covariant}, which intend to develop a path integral quantization of $BF$ gravity and thus supplement the canonical (or loop) approach~\cite{rovelli2004quantum,thiemann2007modern} (see for instance Ref.~\cite{oriti_2009} for a compendium of approaches to quantum gravity).
	
	Although the spinfoam approach is a prominent candidate for a fully diffeomorphism and Lorentz invariant quantization of the gravitational field, the knowledge of the canonical structure of the $BF$-type actions on which it is based can in principle be used to establish connections between its results and those of the loop scenario~\cite{alexandrov2012spin}. This is what motivates our interest in studying the Hamiltonian description of the $BF$ formulations for gravity.
	
	In Ref.~\cite{CelMontcqg2920}, the canonical analyses of two constrained $BF$-type actions for general relativity with the Immirzi parameter~\cite{Immirzicqg1410} were performed in a Lorentz-covariant fashion. Both canonical analyses can be related to one another by a suitable redefinition of the canonical variables employed in them. However, the analysis introduced second-class constraints, which were later solved while nonmanifestly preserving Lorentz invariance in Ref.~\cite{CelMontesRom}, obtaining a canonical formulation that leads to the Ashtekar-Barbero variables~\cite{Barberoprd5110} in the time gauge. The structure of the canonical theory containing second-class constraints is actually quite similar to that obtained~\cite{BarrosESa} for the Holst action~\cite{HolstPRD53}, and in fact, by getting rid of those constraints in a manifestly Lorentz-covariant fashion, the results of both theories agree~\cite{Montesinos1801}.
	
	Recently, it was uncovered that it is possible to perform the canonical analysis of the Holst action in a manifestly Lorentz-covariant way without involving second-class constraints at all~\cite{CelMontesRomNoSC}. We expect something analogous to be true for the $BF$-type counterparts mentioned at the beginning of the previous paragraph because they are classically equivalent to the Holst action. In this paper we show that this is indeed the case, that is, we perform the canonical analysis of $BF$ gravity with the Immirzi parameter plus a cosmological constant while manifestly preserving local Lorentz invariance and avoiding the presence of second-class constraints during the whole process. This is achieved by reexamining the solution of the constraint on the $B$ field in each action principle considered in Ref.~\cite{CelMontcqg2920}. In this way, we establish that the manifestly Lorentz-covariant canonical formulation of general relativity involving only first-class constraints~\cite{Montesinos1801,CelMontesRomNoSC} can also be derived, without introducing second-class constraints in the process, from the formulation of general relativity as a constrained $BF$ theory with Immirzi parameter and a cosmological constant.

	The structure of this paper is as follows. In Sec.~\ref{can} we consider the Capovilla-Montesinos-Prieto-Rojas (CMPR) action~\cite{CMPR2001} plus a cosmological constant and perform its 3+1 decomposition; we then provide an appropriate solution of the constraint on the $B$ field that makes the introduction of second-class constraints unnecessary; the description of the phase space of general relativity in terms of manifestly Lorentz-covariant variables subject to first-class constraints only immediately follows. Afterwards, in Sec.~\ref{altact} we focus on an alternative formulation of $BF$ gravity and follow the same guidelines as before, obtaining identical results. Finally, in Sec.~\ref{Concl} we give some conclusions.
	
	\textit{Notation}.---Let $M$ be a four dimensional Lorentzian or Riemannian manifold. Points of $M$ are labeled with coordinates $x^{\mu}$, where Greek letters $\mu, \nu,\alpha,\ldots$ are spacetime indices. To carry out the canonical analysis, we assume that $M$ can be foliated by spacelike leaves diffeomorphic to $\Sigma$ so that $M$ has the global topology of $\mathbb{R} \times \Sigma$, with $\Sigma$ being an orientable three dimensional spatial manifold without boundary (for simplicity). We use local coordinates $(x^{\mu})=(t,x^{a})$ adapted to this foliation, where $t$  and $x^a$ ($a,b, \ldots=1,2,3$) label points on $\mathbb{R}$ and $\Sigma$, respectively. Internal indices $I,J,\ldots=0,\ldots,3$ are raised and lowered with the metric $(\eta_{IJ}):=\text{diag}(\sigma,1,1,1)$, wherein $\sigma=-1$ ($\sigma=+1$) for Lorentzian (Riemannian) manifolds. We denote by $\mathfrak{so}(3,1)$ or $\mathfrak{so}(4)$ the Lie algebra of the gauge group $SO(3,1)$ or $SO(4)$, correspondingly. In the framework of $BF$ gravity~\cite{cqgrevBF}, the fundamental variables to describe pure gravity are a set of $\mathfrak{so}(3,1)$- or $\mathfrak{so}(4)$-valued 2-forms $B^{IJ} (=-B^{JI})$, an $\mathfrak{so}(3,1)$- or $\mathfrak{so}(4)$-valued connection 1-form $\omega^{IJ}(=-\omega^{JI})$ whose curvature is defined by $F^I{}_J=d\omega^I{}_J+ \omega^I{}_K\wedge \omega^K{}_J$, an internal tensor $\varphi_{IJKL}$ with the index symmetries $\varphi_{IJKL}=-\varphi_{JIKL}=-\varphi_{IJLK}=\varphi_{KLIJ}$, and a 4-form $\mu=\tilde{\mu}d^4x$; $d^4x$ is a shorthand for $dt\wedge dx^1\wedge dx^2\wedge dx^3$, which in Sec. \ref{can} is also denoted $dtd^3x$. The weight of tensor densities is either denoted with a tilde ``$\sim$'' or explicitly mentioned somewhere in the paper. The $SO(3,1)$ [or $SO(4)$] totally antisymmetric tensor $\epsilon_{IJKL}$ is such that $\epsilon_{0123 }=+1$. Likewise, the totally antisymmetric spacetime tensor density of weight +1 ($-1$) is denoted by $\tilde{\eta}^{\mu\nu\alpha\beta}$ ($\underaccent{\tilde}{\eta}_{\mu\nu\alpha\beta}$) and satisfies $\tilde{\eta}^{t123}=+1$ ($\underaccent{\tilde}{\eta}_{t123}=+1$). In addition, we define the three-dimensional Levi-Civita symbol as $\underaccent{\tilde}{\eta}_{abc}:=\underaccent{\tilde}{\eta}_{tabc}$ ($\tilde{\eta}^{abc}:=\tilde{\eta}^{tabc}$). The symmetrizer and antisymmetrizer are defined by $V_{(\alpha\beta)}:=(V_{\alpha\beta}+V_{\beta\alpha})/2$ and $V_{[\alpha\beta]}:=(V_{\alpha\beta}-V_{\beta\alpha})/2$, respectively. In addition, for an antisymmetric quantity $V_{IJ}$ we define its internal dual as $\ast V_{IJ}:=(1/2)\epsilon_{IJKL}V^{KL}$ and also the object $\stackrel{(\gamma)}{V} _{IJ}:=P_{IJKL}V^{KL}$ for
	\begin{equation}
	P_{IJKL}:=\frac12\left(\eta_{IJKL}+\frac{1}{\gamma}\epsilon_{IJKL}\right),\label{P}
	\end{equation}
	where $\eta_{IJKL}:=\eta_{IK}\eta_{JL}-\eta_{IL}\eta_{JK}$ and $\gamma\neq0$ is the Immirzi parameter.\footnote{We assume $\gamma\neq\pm\sqrt{\sigma}$, which means that the self-dual and anti-self-dual sectors are excluded in our approach.} Its inverse is given by
	\begin{equation}
	(P^{-1})^{IJKL}=\frac{\gamma^2}{2(\gamma^2-\sigma)}\left(\eta^{IJKL}-\frac{1}{\gamma}\epsilon^{IJKL}\right).
	\end{equation}
	Both the internal dual ``$\ast$'' and $P_{IJKL}$ define invertible maps from the Lie algebra $\mathfrak{so}(3,1)$ [or $\mathfrak{so}(4)$] on itself. ``$\wedge$'' and ``$d$'' stand for the wedge product and the exterior derivative of differential forms, correspondingly. In a coordinate basis the components of the curvature explicitly take the form
	\begin{equation}
		F_{\mu\nu}{}^I{}_J= \partial_{\mu} \omega_{\nu}{}^I{}_J - \partial_{\nu} \omega_{\mu}{}^I{}_J + \omega_{\mu}{}^I{}_K \omega_{\nu}{}^K{}_J - \omega_{\nu}{}^I{}_K \omega_{\mu}{}^K{}_J.\label{F}
	\end{equation}
	
%%%%%%%%%%%%%%%%%%%%%%%%%%%
\section{The CMPR action with $\Lambda$}\label{can}

In the formalism of $BF$ theories for gravity, the CMPR action with a cosmological constant is given by~\cite{MontesVelprd814}
\begin{eqnarray}
S[\omega,B,\varphi,\mu]=&&\int_M \left[B^{IJ}\wedge F_{IJ}[\omega]-\varphi_{IJKL}B^{IJ}\wedge B^{KL}\right.\notag\\
&&+\mu \left(a_1\tensor{\varphi}{_{IJ}^{IJ}}+a_2\epsilon_{IJKL}\varphi^{IJKL}\right)\notag\\
&&\left.-\Theta\mu+l_1 B_{IJ}\wedge B^{IJ}+l_2 B_{IJ}\wedge *B^{IJ}\right]\!,\label{BFact1}
\end{eqnarray}
which is classically equivalent to the Holst action with a cosmological constant. To simplify the action a bit, let us perform the field redefinition
\begin{equation}\label{redef}
\psi_{IJKL}:=\varphi_{IJKL}-\frac{1}{2}l_1\eta_{IJKL}-\frac{1}{2}l_2\epsilon_{IJKL};
\end{equation}
the action \eqref{BFact1} takes the equivalent form
\begin{eqnarray}
S[\omega,B,\psi,\mu]=&&\int_M \left[B^{IJ}\wedge F_{IJ}[\omega]-\psi_{IJKL}B^{IJ}\wedge B^{KL}\right.\notag\\
&&\left.+\mu \left(a_1\tensor{\psi}{_{IJ}^{IJ}}+a_2\epsilon_{IJKL}\psi^{IJKL}-\lambda\right)\right]\!,\label{BFact2}
\end{eqnarray}
where we have defined
\begin{equation}\label{lambda}
\lambda:=\Theta-6a_1l_1-12\sigma a_2l_2.
\end{equation}

\subsection{3+1 decomposition}
By expressing the 2-forms involved in \eqref{BFact2} as $B^{IJ}=\tensor{B}{_{ta}^{IJ}}dt\wedge dx^a+(1/2)\tensor{B}{_{ab}^{IJ}}dx^a\wedge dx^b$ (with an analogous relation for the curvature $F^{IJ}$), the action reads (recall the all the spatial boundary terms are omitted because $\partial\Sigma=\varnothing$)
\begin{eqnarray}
S=&&\int_{\mathbb{R}\times\Sigma}dt d^3x\biggl[\tilde{\Pi}^{aIJ}\dot{\omega}_{aIJ}+\omega_{tIJ}D_a \tilde{\Pi}^{aIJ}\notag\\
&&+\frac12\tilde{\eta}^{abc}B_{ta}{}^{IJ}F_{bcIJ}-2\psi_{IJKL}B_{ta}{}^{IJ}\tilde{\Pi}^{aKL}\notag\\
&& +\tilde{\mu}\left(a_1\psi^{IJ}{}_{IJ}+a_2\epsilon_{IJKL}\psi^{IJKL}-\lambda\right)\biggr],\label{BFact3} %-\partial_a(\tilde{\Pi}^{aIJ}\omega_{tIJ})
\end{eqnarray}
where the overdot ``$\cdot$'' stands for $\partial/\partial t$, $D_a \tilde{\Pi}^{aIJ} = \partial_a  \tilde{\Pi}^{aIJ} + \omega_a{}^{I}{}_K \tilde{\Pi}^{aKJ} + \omega_a{}^{J}{}_K \tilde{\Pi}^{aIK}$, and we have defined 
\begin{equation}
\tilde{\Pi}^{aIJ}:=\frac12\tilde{\eta}^{abc}B_{bc}{}^{IJ}. \label{defPi}
\end{equation}

The equation of motion for $\psi_{IJKL}$ yields
\begin{equation}
B_{ta}{}^{IJ}\tilde{\Pi}^{aKL}+B_{ta}{}^{KL}\tilde{\Pi}^{aIJ}-\tilde{\mu}\left(\frac{a_1}{2}\eta^{IJKL}+a_2\epsilon^{IJKL}\right)=0.\label{constr}
\end{equation}
Let us define the nonvanishing spacetime volume by
\begin{eqnarray}
\tilde{V}&&:=\frac{1}{24}\tilde{\eta}^{\mu\nu\lambda\rho}\epsilon_{IJKL}B_{\mu\nu}{}^{IJ}B_{\lambda\rho}{}^{KL}\notag\\
&&=\frac{1}{3}\epsilon_{IJKL}B_{ta}{}^{IJ}\tilde{\Pi}^{aKL}\neq 0.\label{volume}
\end{eqnarray}
Assuming $a_2\neq0$ and multiplying Eq.~\eqref{constr} by $\epsilon_{IJKL}$, we obtain
\begin{equation}
\tilde{\mu}=\frac{\sigma}{4a_2}\tilde{V},\label{mu}
\end{equation}
which substituted back into~\eqref{constr} implies
\begin{eqnarray}
&&B_{ta}{}^{IJ}\tilde{\Pi}^{aKL}+B_{ta}{}^{KL}\tilde{\Pi}^{aIJ}\notag\\
&&-\frac{\sigma}{4}\tilde{V}\left(\frac{a_1}{2a_2}\eta^{IJKL}+\epsilon^{IJKL}\right)=0.\label{constr1}
\end{eqnarray}
This expression defines a system of 20 equations for the 36 unknowns $B_{ta}{}^{IJ}$ and $\tilde{\Pi}^{aIJ}$. To solve~\eqref{constr1} means that $B_{ta}{}^{IJ}$ and $\tilde{\Pi}^{aIJ}$ must be parametrized in terms of 16 independent variables. 

%%%%%%%%%%%%%%%%%%%
\subsection{Solution of the constraint}

Now we find a generic expression of $B_{ta}{}^{IJ}$ and $\tilde{\Pi}^{aIJ}$ as functions of 16 variables that solves~\eqref{constr1}. Following the ideas developed in Ref.~\cite{CelMontcqg2920}, we add four variables $\uac{M}$ and $M^a$ to the original set of variables $(B_{ta}{}^{IJ}, \tilde{\Pi}^{aIJ})$, so that the system of equations~\eqref{constr1} is enlarged with the following definitions:
\begin{subequations}
	\begin{eqnarray}
	&& \uac{M}:=\frac{\tilde{V}}{\sqrt{|H|}},\label{lapse}\\
	&& M^a:=-\frac{\sigma}{2}\tilde{\eta}^{abc}\uac{\uac{H}}_{bd}B_{tc}{}^{IJ}\tilde{\Pi}^d{}_{IJ}\label{shift},
	\end{eqnarray}
\end{subequations}
with $\uac{\uac{H}}_{ab}$ being the inverse of $\tilde{\tilde{H}}^{ab}$ given by 
\begin{eqnarray}
&& \tilde{\tilde{H}}^{ab}:=\frac{\sigma}{2}\tilde{\Pi}^{aIJ}\tilde{\Pi}^b{}_{IJ},\label{densmet1}
\end{eqnarray}
and where we have assumed that $H:=\det(\tilde{\tilde{H}}^{ab})\neq 0$. Note that $H$ is a scalar density of weight +4. This assumption is justified because $\tilde{\tilde{H}}^{ab}$ becomes an object proportional to the (inverse of the) spatial metric defined on $\Sigma$, which is assumed to be nondegenerate (as in the usual ADM formalism~\cite{Arnowitt2008}). The system of 20 equations~\eqref{constr1} for 36 unknowns $B_{ta}{}^{IJ}$ and $\tilde{\Pi}^{aIJ}$ is thus equivalent to the system of $20+4=24$ equations~\eqref{constr1},~\eqref{lapse}, and~\eqref{shift} for the $18+18+1+3= 40$ variables $B_{ta}{}^{IJ}$, $\tilde{\Pi}^{aIJ}$, $\uac{M}$, and $M^a$.

Now, we write the system of equations~\eqref{constr1},~\eqref{lapse}, and~\eqref{shift} in an alternative form. This can be done as follows. Using the definition~\eqref{lapse}, Eq.~\eqref{constr1} acquires the form
\begin{eqnarray}
&&B_{ta}{}^{IJ}\tilde{\Pi}^{aKL}+B_{ta}{}^{KL}\tilde{\Pi}^{aIJ}\notag\\
&&-\frac{\sigma}{4}\sqrt{|H|}\uac{M}\left(\frac{a_1}{2a_2}\eta^{IJKL}+\epsilon^{IJKL}\right)=0.\label{constr1.1}
\end{eqnarray}
Multiplying this equation by $\tilde{\Pi}^b{}_{KL}$ and using~\eqref{densmet1}, we obtain
\begin{eqnarray}
&&B_{ta}{}^{IJ}+\frac{\sigma}{2}\uac{\uac{H}}_{ab}B_{tc}{}^{KL}\tilde{\Pi}^b{}_{KL}\tilde{\Pi}^{cIJ}\notag\\
&&-\frac{1}{8a_2}\sqrt{|H|}\uac{M}\uac{\uac{H}}_{ab}(a_1\tilde{\Pi}^{bIJ}+2a_2\ast\tilde{\Pi}^{bIJ})=0.\label{eq1}
\end{eqnarray}
Multiplying the last equation by $\uac{\uac{H}}_{bc}\tilde{\Pi}^c{}_{IJ}$ yields 
\begin{eqnarray}
&&\uac{\uac{H}}_{bc}B_{ta}{}^{IJ}\tilde{\Pi}^c{}_{IJ}+\uac{\uac{H}}_{ac}B_{tb}{}^{IJ}\tilde{\Pi}^c{}_{IJ}\notag\\
&&+\frac{\sigma}{4}\sqrt{|H|}\uac{M}\uac{\uac{H}}_{ac}\uac{\uac{H}}_{bd}\left(\tilde{\tilde{\varphi}}^{cd}-\frac{a_1}{a_2}\tilde{\tilde{H}}^{cd}\right)=0, \label{eq2}
\end{eqnarray}
where we have made the definition
\begin{eqnarray}
&&\tilde{\tilde{\varphi}}^{ab}:=-\sigma\ast\tilde{\Pi}^{aIJ}\tilde{\Pi}^b{}_{IJ}.\label{phi}
\end{eqnarray}
On the other hand, we can rewrite the expression~\eqref{shift} as
\begin{equation}
\uac{\uac{H}}_{bc}B_{ta}{}^{IJ}\tilde{\Pi}^c{}_{IJ}-\uac{\uac{H}}_{ac}B_{tb}{}^{IJ}\tilde{\Pi}^c{}_{IJ}=2\sigma\uac{\eta}_{abc}M^c.\label{eq3}
\end{equation}
Adding together~\eqref{eq2} and \eqref{eq3}, we find
\begin{eqnarray}
&&\uac{\uac{H}}_{ac}B_{tb}{}^{IJ}\tilde{\Pi}^c{}_{IJ}=-\sigma\uac{\eta}_{abc}M^c\notag\\
&&-\frac{\sigma}{8}\sqrt{|H|}\uac{M}\uac{\uac{H}}_{ac}\uac{\uac{H}}_{bd}\left(\tilde{\tilde{\varphi}}^{cd}-\frac{a_1}{a_2}\tilde{\tilde{H}}^{cd}\right).\label{eq4}
\end{eqnarray}
Substituting this expression into the second term of~\eqref{eq1}, we get
\begin{eqnarray}
&&B_{ta}{}^{IJ}=\frac{1}{4}\uac{M}\sqrt{|H|}\uac{\uac{H}}_{ab}\ast\tilde{\Pi}^{bIJ}+\frac{1}{2}\uac{\eta}_{abc}M^c\tilde{\Pi}^{bIJ}\notag\\
&&+\frac{1}{16}\uac{M}\sqrt{|H|}\uac{\uac{H}}_{ac}\uac{\uac{H}}_{bd}\left(\tilde{\tilde{\varphi}}^{cd}+\frac{a_1}{a_2}\tilde{\tilde{H}}^{cd}\right)\tilde{\Pi}^{bIJ}.\label{eq5}
\end{eqnarray}
Substituting this expression for $B_{ta}{}^{IJ}$ into~\eqref{constr1.1}, multiplying the result by $\ast\tilde{\Pi}^e{}_{IJ}\ast\tilde{\Pi}^f{_{KL}}$, and simplifying, we arrive at the equation
\begin{eqnarray}
&&\frac{1}{8}\uac{M}\sqrt{|H|}\left(\uac{\uac{H}}_{bd}\uac{\uac{H}}_{ca}\tilde{\tilde{\varphi}}^{ce}\tilde{\tilde{\varphi}}^{bf}\tilde{\tilde{\varphi}}^{da}+\frac{a_1}{a_2}\uac{\uac{H}}_{bc}\tilde{\tilde{\varphi}}^{be}\tilde{\tilde{\varphi}}^{cf}\right.\notag\\
&&-4\sigma\tilde{\tilde{\varphi}}^{ef}-4\sigma\frac{a_1}{a_2}\tilde{\tilde{H}}^{ef}\biggr)=0.
\end{eqnarray}
Because both $\uac{M}$ and $H$ are nonvanishing, we obtain a cubic equation for $\tilde{\tilde{\varphi}}^{ab}$ that can be rewritten as
\begin{eqnarray}
&&(\tilde{\tilde{\varphi}}^{eb}+2\sqrt{\sigma}\tilde{\tilde{H}}^{eb})\uac{\uac{H}}_{bc}\left(\tilde{\tilde{\varphi}}^{cd} + 
\frac{a_1}{a_2}\tilde{\tilde{H}}^{cd}\right)\notag\\
&&\times\uac{\uac{H}}_{da}(\tilde{\tilde{\varphi}}^{af}-2\sqrt{\sigma}\tilde{\tilde{H}}^{af})=0.
\end{eqnarray}
This equation has three solutions:
\begin{subequations}
	\begin{eqnarray}
	&&\tilde{\tilde{\varphi}}^{ab}=\pm 2\sqrt{\sigma}\tilde{\tilde{H}}^{ab},\label{sol1}\\
	&&\tilde{\tilde{\varphi}}^{ab}=-\frac{a_1}{a_2}\tilde{\tilde{H}}^{ab}.\label{sol2}
	\end{eqnarray}
\end{subequations}
Note that the solutions \eqref{sol1} are associated to the self-dual and anti-self-dual sectors, which will not be considered here. So, we will skip them and focus on the solution \eqref{sol2} only. Using precisely \eqref{sol2}, we see that the last term of \eqref{eq5} vanishes. Therefore, we have shown that the system of $20+4=24$ equations \eqref{constr1},~\eqref{lapse}, and~\eqref{shift} for the $18+18+4=40$ variables $B_{ta}{}^{IJ}$, $\tilde{\Pi}^{aIJ}$, $\uac{M}$, and $M^a$ is equivalent to the following $18+6=24$ equations~\cite{CelMontcqg2920}:
\begin{subequations}
	\begin{eqnarray}
	&&B_{ta}{}^{IJ}=\frac{1}{4}\uac{M}\sqrt{|H|}\uac{\uac{H}}_{ab}\ast\tilde{\Pi}^{bIJ}+\frac{1}{2}\uac{\eta}_{abc}M^c\tilde{\Pi}^{bIJ},\label{solBta}\\
	&&\ast\tilde{\Pi}^{aIJ}\tilde{\Pi}^b{}_{IJ}-\frac{ a_1}{2a_2}\tilde{\Pi}^{aIJ}\tilde{\Pi}^b{}_{IJ}=0, \label{constr2nd}
	\end{eqnarray}
\end{subequations}
for the same variables $B_{ta}{}^{IJ}$, $\tilde{\Pi}^{aIJ}$, $\uac{M}$, and $M^a$. 

It is at this point where the canonical analysis performed here deviates from the canonical analysis reported in Ref.~\cite{CelMontcqg2920}. There, the constraint \eqref{constr2nd} is considered as a primary constraint of the theory and handled according to Dirac's method. In contrast, here we follow a different approach: we explicitly solve~\eqref{constr2nd}. Note that Eq.~\eqref{constr2nd} can be seen as a system of 6 equations for the 18 unknowns $\tilde{\Pi}^{aIJ}$. Therefore, the solution of Eq.~\eqref{constr2nd} must involve 12 independent variables that we denote by $\tilde{\Pi}^{aI}$. The most generic parametrization for $\tilde{\Pi}^{aIJ}$ is 
\begin{equation}
\tilde{\Pi}^{aIJ}=( k_1 \eta^{IJ}{}_{KL}+ k_2 \epsilon^{IJ}{}_{KL})\tilde{\Pi}^{aK}m^L,\label{SolPi}
\end{equation}
where $k_1$ and $k_2$ are constants, whereas $m^I$ is an internal vector constructed out of $\tilde{\Pi}^{aI}$ as
\begin{equation}
m_I=\frac{1}{6\sqrt{h}}\epsilon_{IJKL}\uac{\eta}_{abc}\tilde{\Pi}^{aJ}\tilde{\Pi}^{bK}\tilde{\Pi}^{cL},
\end{equation}
with $h$ of weight $+4$ given by $h:=\det(\tilde{\tilde{h}}^{ab})>0$, for $\tilde{\tilde{h}}^{ab}:=\tilde{\Pi}^{aI} \tilde{\Pi}^b{}_I$; $m^I$ satisfies the orthogonality relation $\tilde{\Pi}^{aI}m_I=0$ and is normalized according to $m_Im^I=\sigma$.

Substituting~\eqref{SolPi} into~\eqref{constr2nd}, we find that the constants $k_1$ and $k_2$ are related to $a_1,\ a_2$ by
\begin{equation}
\frac{a_1}{a_2}=\frac{4 k_1 k_2 }{\sigma (k_1)^ 2+(k_2)^2}. \label{rela1a2}
\end{equation}

It remains to replace~\eqref{SolPi} into~\eqref{solBta} to find $B_{ta}{}^{IJ}$ as a function of $\tilde{\Pi}^{aI}$, $\uac{M}$, and $M^a$. This, together with~\eqref{SolPi} and~\eqref{rela1a2}, is the desired solution of the constraint~\eqref{constr1} because the variables  $B_{ta}{}^{IJ}$ and $\tilde{\Pi}^{aIJ}$ are thus expressed as functions of the $12+1+3=16$ independent variables $\tilde{\Pi}^{aI}$, $\uac{M}$, and $M^a$. Before doing that, we first substitute Eq.~\eqref{SolPi} into Eq.~\eqref{densmet1}, and obtain the relation
\begin{equation}
\tilde{\tilde{H}}^{ab}=\left[(k_1)^2+\sigma (k_2)^2\right]\tilde{\tilde{h}}^{ab},\label{metrich}
\end{equation}
which implies $H=\left[(k_1)^2+\sigma (k_2)^2\right]^3h$; since both $H$ and $h$ are nonvanishing, the condition $(k_1)^2\neq-\sigma (k_2)^2$ must be satisfied.~Combining~this expression with~\eqref{lapse}, Eq.~\eqref{mu} reads $\tilde{\mu}=\sigma\left| (k_1)^2+\sigma (k_2)^2\right|^{3/2}\sqrt{h}\uac{M}/4a_2$. Now, substituting Eq.~\eqref{SolPi} into Eq.~\eqref{solBta}, we get
\begin{eqnarray}
B_{ta}{}^{IJ}=&&-\frac{\sigma}{8}\text{sgn}\left[(k_1)^2+\sigma (k_2)^2\right]\left| (k_1)^2+\sigma (k_2)^2\right|^{1/2}\uac{M}\notag\\
&&\times\uac{\eta}_{abc}\left(k_1 \eta^{IJ}{}_{KL}+ k_2 \epsilon^{IJ}{}_{KL}\right)\tilde{\Pi}^{bK}\tilde{\Pi}^{cL}\notag\\
&&+\frac{1}{2}\uac{\eta}_{abc}M^c\left( k_1 \eta^{IJ}{}_{KL}+ k_2 \epsilon^{IJ}{}_{KL}\right)\tilde{\Pi}^{bK}m^{L},\label{solBta0}
\end{eqnarray}
or, equivalently, 
\begin{eqnarray}
&&B_{ta}{}^{IJ}=-\frac{\sigma}{4|\gamma|}\text{sgn}(\gamma^2+\sigma) k_1 |k_1||\gamma^2+\sigma|^{1/2}\uac{M}\notag\\
&&\times\uac{\eta}_{abc}P^{IJ}{}_{KL}\tilde{\Pi}^{bK}\tilde{\Pi}^{cL}+ k_1 \uac{\eta}_{abc}M^c P^{IJ}{}_{KL}\tilde{\Pi}^{bK}m^L,\label{solBta1}
\end{eqnarray}
where we have identified the Immirzi parameter as $\gamma:=k_1/k_2$. Notice that in the Lorentzian case, the inequality relating the squares of both $k_1$ and $k_2$ implies $\gamma\neq \pm 1$.

Finally, we can verify by direct substitution that~\eqref{SolPi} and \eqref{solBta1} satisfy the constraint \eqref{constr1}, where the relation $\uac{M}=|(k_1)^2+\sigma (k_2)^2|^{-3/2}\tilde{V}/\sqrt{h}$ as well as~\eqref{rela1a2} have to be taken into account during the process.

%%%%%%%%%%%%%%%%
\subsection{Back to the action}

Let us define the (densitized) lapse and shift vector respectively as
\begin{subequations}
	\begin{eqnarray}
	&&\uac{N}:=\frac{1}{4|\gamma|}\text{sgn}(\gamma^2+\sigma)|k_1||\gamma^2+\sigma|^{1/2}\uac{M},\label{lapse1}\\
	&&N^a:=\frac{1}{2}M^a,\label{shift1}
	\end{eqnarray}
\end{subequations}
so that \eqref{solBta1} takes the form
\begin{eqnarray}
B_{ta}{}^{IJ}=&&-\sigma k_1 \uac{N}\uac{\eta}_{abc}P^{IJ}{}_{KL}\tilde{\Pi}^{bK}\tilde{\Pi}^{cL}\notag\\
&&+2 k_1 \uac{\eta}_{abc}N^cP^{IJ}{}_{KL}\tilde{\Pi}^{bK}m^L,\label{solBta1.1}
\end{eqnarray}
Substituting \eqref{SolPi} and \eqref{solBta1.1} into the action \eqref{BFact3}, and doing some algebra, we arrive at
\begin{eqnarray}
S=&&- k_1 \int_{\mathbb{R}\times\Sigma}dt d^3x\Bigl[-2\tilde{\Pi}^{aI}m^J\partial_t\ovg{\omega}_{aIJ}+\omega_{tIJ}\tilde{\mathcal{G}}^{IJ}\notag\\
&&-N^a\tilde{\mathcal{V}}_a-\uac{N}\tilde{\tilde{\mathcal{H}}}\Bigr],\label{BFact4} %\partial_a(2\tilde{\Pi}^{aI}m^J\ovg{\omega}_{tIJ})
\end{eqnarray}
wherein we have defined
\begin{subequations}
	\begin{eqnarray}
	&& \tilde{\mathcal{G}}^{IJ}:=-2\tensor{P}{^{IJ}_{KL}}\left[\partial_a(\tilde{\Pi}^{aK}m^L)+2\tensor{\omega}{_a^{K}_M}\tilde{\Pi}^{a[M}m^{L]}\right],\notag\\
	\label{Gauss}\\
	&& \tilde{\mathcal{V}}_a:=-2\tilde{\Pi}^{bI}m^J\ovg{F}_{abIJ},\label{vect}\\
	&&\tilde{\tilde{\mathcal{H}}}:=-\sigma \tilde{\Pi}^{aI}\tilde{\Pi}^{bJ}\ovg{F}_{abIJ}+2\sigma\Lambda \sqrt{h}.\label{scal}
	\end{eqnarray}
\end{subequations}
Here, we have identified the cosmological constant $\Lambda$ as
\begin{equation}
\Lambda:=-\frac{k_1}{2a_2}\left(1+\frac{\sigma}{\gamma^2}\right)\lambda.\label{cosmocons}
\end{equation}
Notice that \eqref{rela1a2} can be written as
\begin{equation}
\frac{a_1}{a_2}=\frac{4\sigma\gamma}{\gamma^2+\sigma}.
\end{equation}
Using this in~\eqref{cosmocons} and substituting the value~\eqref{lambda} for $\lambda$, we can fix the following value for $\Theta$:
\begin{equation}
\Theta=12\sigma a_2l_2+6a_1l_1-\frac{\sigma\Lambda a_1}{2 k_2}.
\end{equation}
This value agrees with the one obtained in Ref.~\cite{MontesVelprd814}.

The action \eqref{BFact4} corresponds to what was obtained after performing the $3+1$ decomposition of the Holst action in Ref.~\cite{CelMontesRomNoSC}. The next step consists in expressing the 18 components of the connection $\omega_{aIJ}$ in terms of $12+6$ variables $C_{aI}$ and $\uac{\lambda}_{ab}\ (=\uac{\lambda}_{ba})$ as
\begin{equation}\label{connec_par}
\ovg{\omega}_{aIJ}=M_a{}^b{}_{IJK}C_b{}^K+\uac{\lambda}_{ab}\tilde{N}^b{}_{IJ},
\end{equation}
and the canonical analysis follows exactly the same path of Ref.~\cite{CelMontesRomNoSC}. In brief, the variables $C_{aI}$ together with $\tilde{\Pi}^{aI}$ make up the symplectic structure of the theory, whereas the variables $\uac{\lambda}_{ab}$ are found to be auxiliary variables that appear quadratically in the action. After eliminating the latter by using their own equation of motion and later redefining the Lagrange multiplier in front of the Gauss constraint (or proceeding in the reverse order), the action~\eqref{BFact4} takes the form
\begin{eqnarray}
S=&&-k_1\int_{\mathbb{R}\times\Sigma}dtd^3x \left ( 2\tilde{\Pi}^{aI}\dot{C}_{aI}-\lambda_{IJ}\tilde{\mathcal{G}}^{IJ}-2N^a\tilde{\mathcal{D}}_a\right. \notag\\
&& \left.-\uac{N}\tilde{\tilde{\mathcal{H}}} \right ),\label{canonical}
\end{eqnarray}
where $\lambda_{IJ}$, $N^a$, and $\uac{N}$ play the role of Lagrange multipliers imposing the constraints
\begin{subequations}
	\begin{eqnarray}
	&&\tilde{\mathcal{G}}^{IJ}=2\tilde{\Pi}^{a[I}C_a{}^{J]}+4P^{IJ}{}_{KL}\tilde{\Pi}^{a[K}m^{M]}\Gamma_a{}^L{}_M\approx 0,\label{Gauss4}\\
	&&\tilde{\mathcal{D}}_a:=2\tilde{\Pi}^{bI}\partial_{[a}C_{b]I}-C_{aI}\partial_b\tilde{\Pi}^{bI}\approx 0,\label{diff}\\
	&& \tilde{\tilde{\mathcal{H}}}:=-\sigma\tilde{\Pi}^{aI}\tilde{\Pi}^{bJ}R_{abIJ} +2\tilde{\Pi}^{a[I|}\tilde{\Pi}^{b|J]}\Biggl[C_{aI}C_{bJ}\nonumber\\
	&&+2C_{aI}\ovg{\Gamma}_{bJK}m^K+\left(\Gamma_{aIK}+\frac{2}{\gamma}\ast\Gamma_{aIK}\right)\Gamma_{bJL}m^Km^L\nonumber\\
	&&+\frac{1}{\gamma^2}q^{KL}\Gamma_{aIK}\Gamma_{bJL}\Biggr]+2\sigma\Lambda \sqrt{h}\approx 0.\label{scalar3}
	\end{eqnarray}
\end{subequations}
This is the manifestly Lorentz-covariant canonical formulation of the Holst action that was first found in Ref.~\cite{Montesinos1801}. Therefore, the same canonical formulation emerges after performing, in a Lorentz-covariant fashion and completely avoiding the appearance of second-class constraints, the canonical analysis of the CMPR action as explained in this section, which should not come as a surprise since this action and the Holst action share the same classical dynamics.

%%%%%%%%%%%%%%%%%%%%
\section{Alternative action}\label{altact}

Instead of introducing the Immirzi parameter in the constraint on the field $\varphi_{IJKL}$ as in \eqref{BFact1}, let us consider the action that incorporates it as the (inverse of the) coupling constant of a $BF$-type term involving $\ast B$~\cite{MontesVelprd856}, namely,
\begin{eqnarray}
&&S[\omega,B,\varphi,\mu]=\int_M \left[\left(B^{IJ}+\frac{1}{\gamma}\ast B^{IJ}\right)\wedge F_{IJ}[\omega]\right.\notag\\
&&-\varphi_{IJKL}B^{IJ}\wedge B^{KL}+\mu\epsilon_{IJKL}\varphi^{IJKL}\notag\\
&&-\Theta\mu+l_1 B_{IJ}\wedge B^{IJ}+l_2 B_{IJ}\wedge *B^{IJ}\biggr],\label{BFactalt}
\end{eqnarray}
or
\begin{eqnarray}
S[\omega,B,\psi,\mu]=&&\int_M \biggl[\ovg{B}{}^{IJ}\wedge F_{IJ}[\omega]-\psi_{IJKL}B^{IJ}\wedge B^{KL}\notag\\
&&+\mu(\epsilon_{IJKL}\psi^{IJKL}-\bar{\lambda})\biggr],\label{BF2act1}
\end{eqnarray}
where the redefinition \eqref{redef} was used again and
\begin{equation}
\bar{\lambda}:=\Theta-12\sigma l_2.\label{const2}
\end{equation}
We point out that the actions \eqref{BFact1} and \eqref{BFactalt}, or their corresponding ones after the redefinition of the Lagrange multiplier $\varphi_{IJKL}$, can be mapped into each other at the Lagrangian~\cite{MercSIGMA72011} and Hamiltonian~\cite{CelMontcqg2920} levels, although the approach of Ref.~\cite{CelMontcqg2920} involves second-class constraints. By proceeding as in the previous section (and without introducing second-class constraints either), in what follows we show that the canonical analysis of \eqref{BF2act1} leads to the same structural form of the action \eqref{canonical}.

%%%%%%%%%%%%%%%%
\subsection{3+1 decomposition}
The $3+1$ decomposition of the action \eqref{BF2act1} yields
\begin{eqnarray}
&&S=\int_{\mathbb{R}\times\Sigma}dt d^3x\left[\ovg{\tilde{\Pi}}{}^{aIJ}\dot{\omega}_{aIJ}+\omega_{tIJ} D_a \ovg{\tilde{\Pi}}{}^{aIJ}\right.\notag\\
&& +\frac12\tilde{\eta}^{abc}B_{ta}{}^{IJ}\ovg{F}_{bcIJ}-2\psi_{IJKL}B_{ta}{}^{IJ}\tilde{\Pi}^{aKL}\notag\\
&&+\tilde{\mu}\left(\epsilon_{IJKL}\psi^{IJKL}-\bar{\lambda}\right)\biggr],\label{BF2act3}%-\partial_a(\ovg{\tilde{\Pi}}{}^{aIJ}\omega_{tIJ})+
\end{eqnarray}
where $\tilde{\Pi}^{aIJ}$ is given by~\eqref{defPi} and $D_a \ovg{\tilde{\Pi}}{}^{aIJ} = \partial_a \ovg{\tilde{\Pi}}{}^{aIJ} + \omega_a{}^I{}_K 
\ovg{\tilde{\Pi}}{}^{aKJ} + \omega_a{}^J{}_K 
\ovg{\tilde{\Pi}}{}^{aIK}$. The constraint imposed by $\psi_{IJKL}$ is
\begin{equation}
B_{ta}{}^{IJ}\tilde{\Pi}^{aKL}+B_{ta}{}^{KL}\tilde{\Pi}^{aIJ}-\tilde{\mu}\epsilon^{IJKL}=0.\label{constr3}
\end{equation}
Multiplying this expression by $\epsilon_{IJKL}$ allows us to fix the value for $\tilde{\mu}$ as
\begin{equation}
\tilde{\mu}=\frac{\sigma}{4}\tilde{V},\label{volume2}
\end{equation}
with $\tilde{V}$ still defined as in~\eqref{volume}. So, the constraint \eqref{constr3} reads
\begin{equation}
B_{ta}{}^{IJ}\tilde{\Pi}^{aKL}+B_{ta}{}^{KL}\tilde{\Pi}^{aIJ}-\frac{\sigma}{4}\tilde{V}\epsilon^{IJKL}=0,\label{constr4}
\end{equation}
which corresponds to the constraint \eqref{constr1} with $a_1=0$ and $a_2=1$~\cite{kras2009-26-5}. We will take advantage of the results of Sec.~\ref{can}. Substituting $a_1=0$ in~\eqref{rela1a2} implies either $k_1\neq 0$ and $k_2=0$, or $k_1=0$ and $k_2\neq 0$. Let us analyze these cases separately.

%%%%%%%%%%%%%%%%%%%%%%
\subsection{Case $k_1\neq 0$ and $k_2=0$}
This choice implies, from~\eqref{SolPi}, \eqref{solBta0}, \eqref{lapse1}, and \eqref{shift1}, that
\begin{subequations}
	\begin{eqnarray}
	&&\hspace{-5mm}\tilde{\Pi}^{aIJ}=2k_1\tilde{\Pi}^{a[I}m^{J]},\label{solPi2}\\
	&&\hspace{-5mm}B_{ta}{}^{IJ}=-\sigma k_1\uac{N}\uac{\eta}_{abc}\tilde{\Pi}^{bI}\tilde{\Pi}^{cJ}+2k_1\uac{\eta}_{abc}N^c\tilde{\Pi}^{b[I}m^{J]},\label{solBta2}
	\end{eqnarray}
\end{subequations}
for $\uac{N}=|k_1|\uac{M}/4$. Using this expression, the value for $\tilde{\mu}$ is, from~\eqref{lapse}, \eqref{metrich}, and \eqref{volume2}, $\tilde{\mu}=\sigma (k_1)^2\sqrt{h}\uac{N}$. Substituting all this into the action \eqref{BF2act3}, it acquires the form
\begin{eqnarray}
S=&&-k_1\int_{\mathbb{R}\times\Sigma}dt d^3x\biggl[-2\tilde{\Pi}^{aI}m^J\partial_t\ovg{\omega}_{aIJ}\notag\\
&&+\omega_{tIJ}\tilde{\mathcal{G}}^{IJ}-N^a\tilde{\mathcal{V}}_a-\uac{N}\tilde{\tilde{\mathcal{H}}}\biggr],\label{BFact5}%\partial_a(2\tilde{\Pi}^{aI}m^J\ovg{\omega}_{tIJ})
\end{eqnarray}
with exactly the same expressions \eqref{Gauss}--\eqref{scal} for $\tilde{\mathcal{G}}^{IJ}$, $\tilde{\mathcal{V}}_a$, and $\tilde{\tilde{\mathcal{H}}}$. In this case we have identified the cosmological constant $\Lambda$ as $\Lambda=-k_1\bar{\lambda}/2$, which using~\eqref{const2} yields
\begin{equation}
\Theta=-2\frac{\Lambda}{k_1}+12\sigma l_2.
\end{equation}
This value agrees with the one found in Ref.~\cite{MontesVelprd856} for the solution $B^{IJ}=\kappa_1\ast(e^I\wedge e^J)$, with $\kappa_1=-k_1$. Notice that the action \eqref{BFact5} looks exactly like the action \eqref{BFact4}, the only difference being that whereas in the latter the Immirzi parameter arises as the quotient of $k_1$ and $k_2$ ($\gamma\neq\pm 1$ in the Lorentzian case), in the former the Immirzi parameter and the constant $k_1$ are not related to one another (and the value $\gamma=\pm1$ in the Lorentzian case is not forbidden). In consequence, the parametrization \eqref{solPi2}--\eqref{solBta2} leads to the same results of the CMPR action (as long as the identifications of the lapse function, shift vector, and cosmological constant are the ones indicated in each case).

%%%%%%%%%%%%%%%%%%%%%%
\subsection{Case $k_1=0$ and $k_2\neq 0$}
This choice implies, from \eqref{SolPi}, \eqref{solBta0}, and \eqref{shift1}, that
\begin{subequations}
	\begin{eqnarray}
	&&\tilde{\Pi}^{aIJ}=k_2\epsilon^{IJ}{}_{KL}\tilde{\Pi}^{aK}m^{L},\label{solPi3}\\
	&&B_{ta}{}^{IJ}=-\frac{\sigma}{2} k_2\uac{N}\uac{\eta}_{abc}\epsilon^{IJ}{}_{KL}\tilde{\Pi}^{bK}\tilde{\Pi}^{cL}\notag\\
	&&\hspace{14mm}+k_2\uac{\eta}_{abc}N^c\epsilon^{IJ}{}_{KL}\tilde{\Pi}^{bK}m^{L},\label{solBta3}
	\end{eqnarray}
\end{subequations}
for $\uac{N}=\sigma|k_2|\uac{M}/4$ from \eqref{lapse1}. Using this value, the expression for $\tilde{\mu}$ is, from \eqref{lapse}, \eqref{metrich}, and \eqref{volume2}, $\tilde{\mu}=(k_2)^2\sqrt{h}\uac{N}$. Using all this information, the action \eqref{BF2act3} can be rewritten as
\begin{eqnarray}
&&S=k_2\int_{\mathbb{R}\times\Sigma}dt d^3x\biggl\{-2\tilde{\Pi}^{aI}m^J\partial_t\ast\ovg{\omega}_{aIJ}+2\ast\ovg{\omega}{}_{tIJ}\notag\\
&&\times\Bigl[\partial_a\left(\tilde{\Pi}^{aI}m^J\right)+2\omega_a{}^I{}_M\tilde{\Pi}^{a[M}m^{J]}\Bigr]-2N^a\tilde{\Pi}^{bI}m^J\notag\\
&&\times\ast\ovg{F}_{abIJ}-\sigma \uac{N}\Bigl(\tilde{\Pi}^{aI}\tilde{\Pi}^{bJ}\ast\ovg{F}_{abIJ}+\sigma k_2 \bar{\lambda} \sqrt{h}\Bigr)\biggr\} .\label{BFact6}%-\partial_a(2D\tilde{\Pi}^{aI}m^J\ast\ovg{\omega}_{tIJ})
\end{eqnarray}
Notice however that
\begin{subequations}
	\begin{eqnarray}
	&&\ast\ovg{V}_{IJ}=\frac{\sigma}{\gamma}Q_{IJKL}V^{KL},\\
	&& Q_{IJKL}:=\frac12(\eta_{IJKL}+\sigma\gamma\epsilon_{IJKL}).\label{Q}
	\end{eqnarray}
\end{subequations}
$Q_{IJKL}$ is just the internal tensor $P_{IJKL}$ with $\gamma$ replaced with $\sigma\gamma^{-1}$. Using \eqref{Q} in the action \eqref{BFact6}, we finally arrive at
\begin{eqnarray}
S=&&-\frac{\sigma k_2}{\gamma}\int_{\mathbb{R}\times\Sigma}dt d^3x\Bigl[-2\tilde{\Pi}^{aI}m^J\partial_t(Q_{IJKL}\omega_a{}^{KL})\notag\\
&&+\omega_{tIJ}\tilde{\mathcal{G}}^{IJ}-N^a\tilde{\mathcal{V}}_a-\uac{N}\tilde{\tilde{\mathcal{H}}}\Bigr],\label{BFact7}%\partial_a(2\tilde{\Pi}^{aI}m^JQ_{IJKL}\tensor{\omega}{_t^{KL}})
\end{eqnarray}
for
\begin{subequations}
	\begin{eqnarray}
	&& \tilde{\mathcal{G}}^{IJ}:=-2Q^{IJ}{}_{KL}\left[\partial_a\left(\tilde{\Pi}^{aK}m^L\right)+2\omega_a{}^K{}_M\tilde{\Pi}^{a[M}m^{L]}\right],\notag\\\label{Gauss3}\\
	&& \tilde{\mathcal{V}}_a:=-2\tilde{\Pi}^{bI}m^J Q_{IJKL}F_{ab}{}^{KL},\label{vect3}\\
	&&\tilde{\tilde{\mathcal{H}}}:=-\sigma \tilde{\Pi}^{aI}\tilde{\Pi}^{bJ}Q_{IJKL}F_{ab}{}^{KL}+2\sigma\Lambda \sqrt{h},\label{scal3}
	\end{eqnarray}
\end{subequations}
where we have identified the cosmological constant $\Lambda$ as $\Lambda=-\gamma k_2 \bar{\lambda}/2$, which in turn implies, from \eqref{const2}, that
\begin{equation}
\Theta=12\sigma l_2-\frac{2\Lambda}{\gamma k_2}.
\end{equation}
This value agrees with the one found in Ref.~\cite{MontesVelprd856} for the solution $B^{IJ}=\kappa_2e^I\wedge e^J$, with $\kappa_2=-\sigma k_2$. The action \eqref{BFact7} is, up to a constant global factor, the same action \eqref{BFact4} or \eqref{BFact5} with the Immirzi parameter replaced with $\sigma\gamma^{-1}$. Taking this into account, we can proceed as we did at the end of Sec. \ref{can} and finish up with an action principle exactly as \eqref{canonical} with the simultaneous replacements $k_1\rightarrow\sigma k_2/\gamma$ and $\gamma\rightarrow\sigma\gamma^{-1}$. Thus, as expected, the resulting canonical theory contains the same physics as the action \eqref{canonical}.

\section{Conclusions}\label{Concl}

In this paper we have performed, in a manifestly Lorentz-covariant fashion and without introducing second-class constraints, the canonical analysis of $BF$ gravity with the Immirzi parameter and a cosmological constant. We consider the CMPR action in Sec.~\ref{can} and an alternative form of the action in Sec.~\ref{altact}, whose relation has already been established at Lagrangian~\cite{MercSIGMA72011} and Hamiltonian~\cite{CelMontcqg2920} frameworks, although involving second-class constraints in the latter case. The strategy we followed in both cases consisted in first performing the $3+1$ decomposition of the action principles and then appropriately handling the constraint imposed by the Lagrange multiplier $\psi_{IJKL}$ on the $B$ field to make contact with the homologous canonical analysis of the Holst action reported in Ref.~\cite{CelMontesRomNoSC}. Since the constraint for the action of Sec.~\ref{altact} is a particular case of the constraint for the CMPR action, here we focus on discussing the latter.  With the help of the introduction of \eqref{lapse} and \eqref{shift}, we can manipulate the aforementioned constraint by breaking it down into a set restraining only the magnetic components $\tilde{\Pi}^{aIJ}\sim B_{ab}{}^{IJ}$ according to~\eqref{constr2nd}, whereas the remaining set of constraints is used to fix the electric components $B_{ta}{}^{IJ}$ as in \eqref{solBta}. This splitting was already noted in Ref.~\cite{CelMontcqg2920}, where the constraint~\eqref{constr2nd} was treated as a primary constraint of the theory; its time evolution then generated a secondary constraint, and both constraints turned out to be second class. 

Here we take a different approach. We explicitly solve~\eqref{constr2nd} and obtain a parametrization of the $B$ field in terms of the 16 independent variables $\tilde{\Pi}^{aI}$, $\uac{M}$, and $M^a$ as depicted in \eqref{SolPi}, \eqref{rela1a2}, and \eqref{solBta0}. Therefore, this parametrization provides a faithful solution of the constraint \eqref{constr1}, which defines a set of 20 equations for the 36 unknowns contained in the $B$ field. With this at hand, the action takes the form \eqref{BFact4}, which exactly matches an intermediate step in the canonical analysis of the Holst action as developed in Ref.~\cite{CelMontesRomNoSC}. From this point on the canonical analysis is entirely the same as that of Ref.~\cite{CelMontesRomNoSC}, resulting in the manifestly Lorentz-covariant formulation embodied in \eqref{canonical}, whose relation to other canonical formulations of general relativity was established in Refs.~\cite{Montesinos1801,CelMontesRomNoSC} (see also Ref.~\cite{MontRomEscCel}). It is worth stressing that no second-class constraints are introduced during the entire process.

Regarding the alternative action of Sec.~\ref{altact}, similar conclusions can be drawn for it as well. In this case there are two independent solutions for the constraint on $\tilde{\Pi}^{aIJ}$, as explained in that section. One of them leads to the same results discussed in the previous paragraph, and the other leads to the same results modulo a global constant factor in the action and a redefinition $\gamma\rightarrow\sigma\gamma^{-1}$ of the Immirzi parameter. Analogous results are actually found at the Lagrangian level in regard to the relation between the alternative $BF$-type action principle and the Holst action with a cosmological constant~\cite{MontesVelprd856}.

For the sake of completeness, it would be interesting to extend our results to the formulation of higher dimensional general relativity as a constrained $BF$ theory~\cite{Freid_Puz} (see also Ref.~\cite{cqgrevBF}) to contrast with those of Ref.~\cite{Bodend_2013}, where the canonical analysis---involving second-class constraints---of the action of Ref.~\cite{Freid_Puz} is sketched. In addition, we expect to make contact with the analysis carried out in Ref.~\cite{Montesinosprog1} for the $n$-dimensional Palatini action. Work on this last issue is in progress.

Although the canonical theory contained in this paper keeps Lorentz invariance manifest, a reinterpretation of it in terms of more canonically quantizable variables such as Lorentz-covariant connections, remains an open question. However, if such a thing were possible, this would enormously contribute not only to getting a canonical quantum theory of gravity with full Lorentz invariance but also to better understand the deep relationship existing between the canonical approach and the spinfoam models for gravity.

%%%%%%%%%%%%%%%%%%%%
\acknowledgments

We thank Diego Gonzalez for carefully reading the manuscript and for his valuable comments. This work was partially supported by Fondo SEP-Cinvestav and by Consejo Nacional de Ciencia y Tecnolog\'{i}a (CONACyT), M\'{e}xico, Grant No.~A1-S-7701. M.~C. gratefully acknowledges the support of a DGAPA-UNAM postdoctoral fellowship.

 	\bibliographystyle{apsrev4-1}
	\bibliography{references}

\end{document}